\documentclass[twocolumn,pra,showpacs,preprintnumbers,amsmath,amssymb]{revtex4-1}
\usepackage{graphicx}
\usepackage{dcolumn}
\usepackage{bm}
\usepackage{amsmath}
\usepackage{hyperref}
\hypersetup{pdfborder = {0 0 0}}

\begin{document}

\title{Experimental and theoretical investigation of 
statistical regimes  in random laser emission}

\author{Emilio Ignesti$^{1}$, Federico Tommasi$^{1*}$, 
Lorenzo Fini$^{1,2}$, Stefano Lepri$^{3}$, Vivekananthan Radhalakshmi$^{2}$, Diederik Wiersma$^{2,4}$ and Stefano Cavalieri$^{1,2}$}

\affiliation{$^{1}$Dipartimento di Fisica e Astronomia, Universit$\grave{a}$
di Firenze, Via Giovanni Sansone 1 I-50019 Sesto Fiorentino, Italy.\\
$^{2}$European Laboratory for Non-Linear Spectroscopy (LENS)\\
 Universit$\grave{a}$ di Firenze, Via Nello Carrara 1, 
 I-50019 Sesto Fiorentino, Italy.\\
 $^{3}$Istituto dei Sistemi Complessi, Consiglio Nazionale delle Ricerche,
 via Madonna del Piano 10, I-50019 Sesto Fiorentino, Italy\\
 $^{4}$ INO-CNR, Largo Fermi 6, I-50125 Firenze, Italy\\
$^{*}$Corresponding author: federico.tommasi@unifi.it\\
}

\date{\today}

\begin{abstract}

We present a theoretical and experimental study aimed at characterizing
 statistical regimes in a random laser. Both the theoretical simulations 
and the experimental results show the possibility of three
region of fluctuations increasing the pumping energy. 
An initial Gaussian regime is followed by a L\'evy statistics and Gaussian
 statistic is recovered again for high pump pulse energy. These different
statistical regimes are possible in a weakly diffusive active medium, 
while the region of L\'evy statistics
disappears when the medium is strongly diffusive 
presenting always a Gaussian regime with smooth
emission spectrum. Experiments and theory agree in identification 
of the key parameters determining the
statistical regimes of the random laser.

\end{abstract}


\maketitle

\section{Introduction}\label{Introduction}

In a random laser, as theorized for the first time by Letokhov in the 1960s 
\cite{Letokhov}, amplification via stimulated emission is provided by multiple
 scattering in a disordered gain medium \cite{Wiersma1,Wiersma2,Cao1}.
In contrast to conventional laser modes, which are determined by the
characteristics of the optical cavity, in a random laser light propagates 
as a random walk through a diffusing material. If the gain experienced along 
the path inside the active medium overcomes the losses, laser emission can occur.

Despite the large number of experimental studies on the emission
 characteristics of different random laser materials as well as the large
 amount of theoretical models reported in literature, a clear
connection between experiment and theory is still lacking to date. 
Because the random nature of light
propagation, with a large number of modes competing for the available gain,
 the emission spectrum and
the spatial properties of the output are in general hard to predict;
 the concept of  ``mode'' itself can
refer to different definitions or interpretations. Moreover,
 together  with narrowing of the
emission spectrum above laser threshold, narrow spikes at random frequencies
 were also reported in some
experimental works.  The presence of these narrow spikes in emission spectra
 led many researchers to
suggest different theoretical possible explanations. For instance,
 Anderson localization was invoked for
a material where the scattering mean free path $\ell_{MFP}$ becomes smaller
 than the radiation
wavelength $\lambda$ \cite{Jiang,Pradhan,Cao2,Vanneste}. Other models were
 proposed in order to explain
the presence of narrow spikes in a weakly diffusive regime 
($\lambda\ll \ell_{MFP}$), where the Anderson
localization constrain is not satisfied: waveguide structure in a material
 with random variable
refractive index \cite{Apalkov}, Fabry-Perot type cavity  determined by the
 geometry of the active
volume within the sample \cite{Wu}, single frequency eigenmodes with a long
 decay time \cite{Molen} and
lasing mode consisting of traveling waves extended through the whole 
system \cite{Vanneste2007}. A recent
work \cite{Ramy} suggests that the crossover from smooth to spiky  spectrum,
 in a weakly scattering medium, can be
interpreted in terms of modes that become uncoupled from mode competition.
 Some experimental
and theoretical studies suggest that random fluctuations in these spectra
 may be due to rare extended
modes, identified as  possible long lifetime random paths within the active
 medium; these so-called
\emph{lucky photons}  are able to experience a very large gain, giving rise
 to spikes in total intensity
spectrum \cite{Mujandar,Mujandar2}. It has also been reported that,  in the presence
 of strong fluctuations,
 the spikes' intensity distribution follows a L\'evy-like
 behavior, i.e.\ a
distribution  that exhibits a \emph{fat tail} associated to infrequent but
 large events \cite{Sharma}.
As a results emission can display a large statistical variability even
 under well controlled
experimental conditions. 

In a previous theoretical work \cite{Lepri}, based on a phase-insensitive intensity
 feedback model, it
was demonstrated that the fluctuations of the emitted light can change,
 upon changing of
a control parameter, from Gaussian to L\'evy
statistical regimes, the latter being characterized by a power-law tail
 in the 
intensity distributions. Actually, the L\'evy regime is limited to a
specific range of gain length, while Gaussian statistics occur with
 both low and high
pumping energy.  Three different statistical regimes are then
 possible depending on the gain
\cite{Lepri}.  Such regimes have been experimentally identified in weakly
scattering media with a short-time pump pulse duration ($\sim$ 30 ps) \cite{Uppu}.
 A more recent work
presents  a theoretical to experimental convergence in
 determination of the first
transition in the statistical regimes as a function of pumping energy
 for different scatterers
concentrations \cite{Uppu2}.  A related experimental investigation
 of L\'evy statistical fluctuations
for  samples with immobile light scatterers can be found in  Ref.~\cite{Zhu2012}.

A short excitation time has been often considered a requirement
 for observation of spiky spectra. A long duration may
 lead to an averaging mechanism
 among modes, thus reducing the fluctuations.

In this work we show experimental evidence of different statistical regimes
 of fluctuations and
different spectra, presenting smooth or spiky structure, as a function of the
 pumping energy in a
nanosecond excitation. We present also a Monte Carlo simulation of 
a two-dimensional (2D) model,  that includes the medium's
population and parallel processing of a large number of random walkers,
 in either a long or short
excitation pulse. In both experiments and numerical simulations,
 three different statistical
regimes by increasing the pumping energy are possible: Gaussian
 transition to L\'evy regime
 and back again for high
pumping energy. We show that the value of the scattering mean free path plays
 a key role in determining the
transition through the different statistical regimes or even in closing
 the region of L\'evy statistics
both in theoretical simulations and experimental results. We then present
 a clear quantitative link
between experiment and theory.

The organization of the paper is the following: In Sec.
 \ref{Theoretical Model} we describe the theoretical model and the structure
 of the Monte Carlo simulation and results are discussed in Sec.
 \ref{Numerical Statistical Analysis}. In Sec.
 \ref{Experimental Set Up and Samples} we describe the experimental
 set up and the measurement method designed and performed in order
 to optimize spectra analysis and statistics of fluctuations.
 The results obtained are presented in Sec. \ref{Experimental Results}.
 Finally,  Sec.  \ref{Conclusion} summarizes and comments on the experimental results and theoretical analysis.

\section{Theoretical Model}\label{Theoretical Model}

The numerical model used in this work is based on an intensity
 feedback that does not include the
effects of the field interference and thus 
treats light propagation through a Monte-Carlo dynamics
 \cite{Mujandar,Lepri}. 
The sample consists of a two-dimensional square region
partitioned in equal square cells  of linear size $\ell$ given
 by the distance covered by light in a
time step $dt$.  
Each cell is labeled by the vector index $(x,y)$, with $x,y$ integers 
$-L/2 \le x\le L/2$, $-L/2 \le y\le L/2$ so that $L^2$ is the
 total number of cells.  The
amplifying medium is modeled by a four-level system  and
 a population of excited atoms, providing
the population inversion $N(x,y)$. 
The spontaneous emission is reproduced by a spectrum of 1001
 arbitrary channels
(labeled from $-$500 to 500 channel) centered on the resonance
 of the transition (channel 0) with a linewidth of $w$ channels. 
The light dynamics is described in terms of a large number
 of random walkers each carrying an energy 
 (or number of photons) $n_i$ with an 
assigned frequency channel $\omega_i$ around the 
central frequency $\omega_0$.  

The simulation consists in a loop of three distinct steps for each $dt$:

\begin{itemize}
\item[(1)] \emph{Spontaneous emission}\\
For each lattice cell, a spontaneous emission event can randomly occur 
with a probability given by $\gamma_0 N dt$. 
If this event occurs, the local population $N$ of the cell is reduced by 
a unity and a new walker is created with an initial energy $n_i=1$, 
a random initial direction of motion 
and a frequency $\omega_i$. The latter is drawn at random 
from a Cauchy distribution of width $w$ centered at the resonance of 
the transition [see Eq.(\ref{stim3} below].     
\item[(2)] \emph{Diffusion}\\
Each walker present in the slab can undergo a scattering event with a
 probability  $\mathcal{P}_s$, that randomly changes its direction. 
The position of each walker is thus updated according the
 trajectory of its motion. 
If a walker reaches a slab boundary it is destroyed and its number
 of photons and frequency 
are recorded for the statistical analysis.
\item[(3)] \emph{Stimulated emission}\\
The population of each lattice cell and the number of photons
 carried by each walker are deterministically updated
according to the following rules:
\begin{eqnarray}
&& n_i  \longrightarrow  [1+\gamma (\omega_i )Ndt]n_i \\
 \label{stim1}
&& N \longrightarrow [1-\gamma(\omega_i)n_idt]N
\label{stim2}
 \end{eqnarray}
where $N$ is population of the lattice cell where the
 $i$-th walker is localized, 
whereas the stimulated emission coefficient $\gamma$
 depends on frequency:
\begin{equation} 
 \gamma(\omega_i)=\frac{\gamma_0}{1+(\omega_i/w)^2}
 \label{stim3}
 \end{equation}
 
\end{itemize}

It should be emphasized that, although each walker evolves independently
from all the others, they all interact with the same population distribution, 
which, in turn, determines the photon number distributions. 
It is thus mandatory to evolve in parallel the 
trajectory of all the walkers that are present in the simulation box. 
This requirement is computationally demanding but is necessary 
to take into account the effect of gain saturation and spatio-temporal coupling 
among different walkers in a self-consistent way. Note also that 
the gain length (i.e.\ the typical length needed to increase the walker energy 
by a factor $e$) is not constant being a function of both time and space. 

In the following, the initial population distribution $N(x,y)$ at $t=0$ 
is assumed to have a Gaussian spatial shape 
\begin{equation}
N(x,y,t=0) =N_o(x,y)=  A \exp\left( - \frac{x^2+y^2}{2\sigma^2}\right) 
\label{gaussian}
\end{equation}
The normalization factor $A$ is assigned to have a prescribed initial
 energy $E_{tot}$: 

\begin{equation}
E_{\text{tot}} =  \hbar\omega_0\int N_o(x,y)dxdy 
\label{gaussia2}
\end{equation}

 This assignment corresponds to considering an infinitely short excitation
 during which the sample absorbs and energy $E_{tot}$ from the pump beam.
 In the following, we also investigate the case in which pumping
 occurs on a finite time $T_p$; in such a case the updating
 of the population follow this rule:

\begin{equation}
N(x,y,t)\longrightarrow N(x,y,t)+R_p(x,y,t)dt 
\label{gaussia3}
\end{equation}
where the \emph{pumping rate} $R_p(x,y,t)$ is given by
    \begin{equation}
    R_p(x,y,t)=
    \begin{cases}
\frac{N_o(x,y)}{T_p} & \text{if $0\le t \le T_p$} \\
0 & \text{if $t > T_p$}

 \end{cases}
\label{gaussia4}
\end{equation}

Thus, the important parameters to be set for each Monte Carlo are: the
spontaneous emission rate $\gamma_0$, the linewidth of the transition $w$,
  and the scattering
probability per unit time $\mathcal{P}_s$. The latter determines
 the actual mean free path 
$\ell_{MFP}$. 
This  that can be compared with the linear size $\sigma$ 
of the extension of the initially excited population, 
as defined by Eq.(\ref{gaussian}).  
It should be noted that $\sigma$ gives a measure of the effective
 size of the 
gain volume, i.e.\ the actual region where amplification occurs.\\
In numerical simulations spatial lengths, time intervals and 
frequencies are expressed in units of $\ell$, $dt$ and
 $dt^{-1}$ respectively, while the energy is reported as the
  number of initial excited molecules in the lattice and in units 
  of $\hbar\omega_0$, where $\omega_0$ is the central 
  frequency of the transition. 

In the following we fix $\sigma$ = 40 and $L$ = 150. We will study
 the statistic regimes 
as a function of other parameters and in particular of $\ell_{MFP}$.

\section{Numerical Statistical Analysis}
\label{Numerical Statistical Analysis}

The central limit theorem  states that the sum of a large number
 of independent, uniformly distributed
random variables tends  to a normal distribution; nonetheless
 there are many systems and observations
for which a non-Gaussian stable model is to be expected.
 In these cases, a distribution with infinite
variance can be described, according to the generalized central limit 
theorem, by a L\'evy-stable
distribution, that takes into account extreme changes in a variables,
 associated with strong
fluctuations with  rare and very large values having an infinite
 variance \cite{alpha0,alpha00}. 
Such distributions are parametrized by the exponent $\alpha$
 (the index of
stability or tail index) that describes the rate at which the
 tail of the distribution tapers 
off:  At $\alpha<2$ the tail exhibits a asymptotically  power-law behavior.
 Upon approaching 
$\alpha = 2$, the distribution tends to a Gaussian one.

In the following we characterize the measured
 data by fitting them to 
a L\'evy-stable distribution. Before entering in the presentation
 of the results,
we emphasize that
this procedure deserves a word of caution. Due to the finite number
 of measurements
it is in general rather difficult to reliably resolve the tail
 of the distributions.
In particular, for marginal cases where $\alpha$ is close to 2 there
 is a certain degree 
of arbitrariness in discriminating between a Gaussian distribution
  and a L\'evy one.
We thus decided to fix an arbitrary threshold value, $\alpha=1.8$, 
above which the characteristics of the intensity distribution
 and emission spectra converge toward a Gaussian behavior leading
 to an attribution of Gaussian regime. 

\begin{figure}[h!]
 \centerline{\scalebox{0.45}{\includegraphics{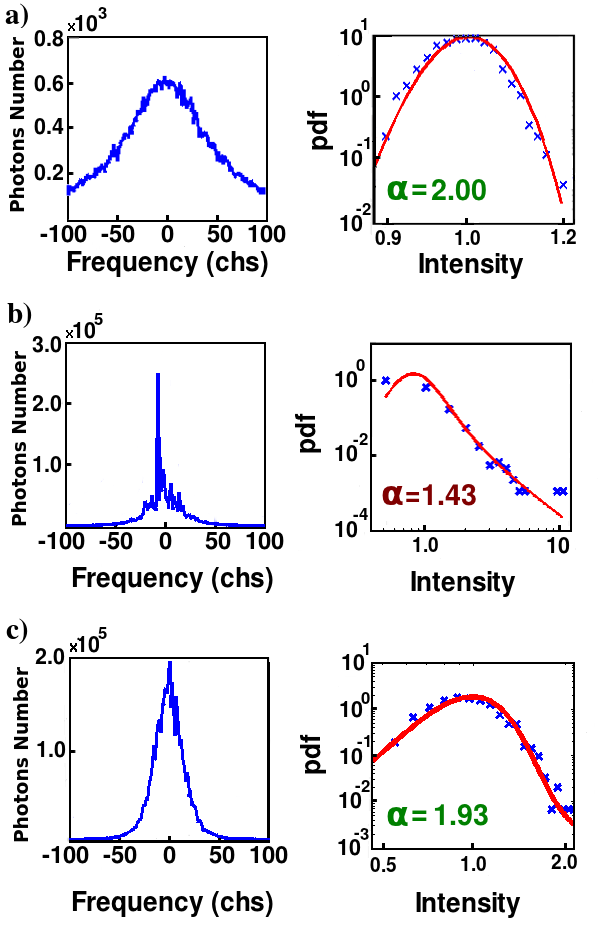}}}
\caption{(Color online) Numerical results for a square lattice 
for three
 different pumping energy for impulsive pumping ($T_p=0$),
 $\ell_{MFP}=40$ and $L=150$. The spectra of emission are expressed in
channels (chs) around the resonance of the transition (chs
=0). The values of energy in 
 units of $\hbar\omega_0$
 [$1.13\times 10^5$ for
  case (a), $3.30\times 10^6$ for (b) 
  and $6.75\times 10^6$ for (c) ] 
 are relative to the condition of sub threshold, near threshold and far 
 above threshold.
 On the left column typical spectra are shown while in the 
 right column
 spectral peak intensity histograms and relative 
 probability density 
 function (pdf) L\'evy-stable fits are
 reported with the value of $\alpha$ parameter.
  Intensity distributions
 are normalized to median and reported in logarithmic scale.
  In cases (a) and (c) the statistical regime is Gaussian, 
  while the case (b) it falls into the L\'evy fluctuations regime.}
 \label{lalande}
\end{figure}

\begin{figure}[h!]
 \centerline{\scalebox{0.45}{\includegraphics{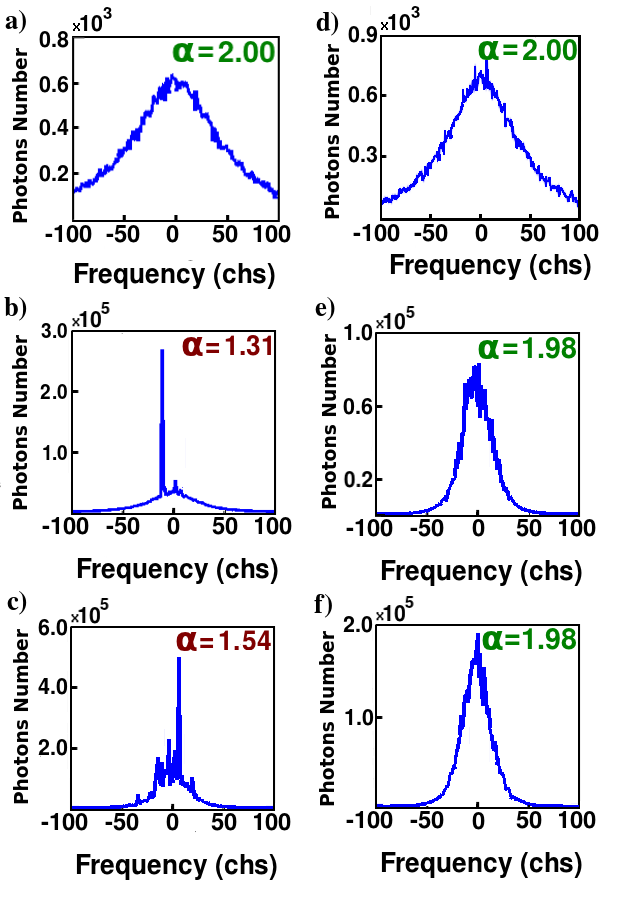}}}
\caption{(Color online) Typical numerical spectra for the
 same pumping energies as in Fig. \ref{lalande} [ $1.13\times 10^5$
  for cases (a) and (d),
  $3.38\times 10^6$ for cases (b) and (e) and $1.13\times 10^5$
   for cases (c) and (f) and $6.75\times 10^6$ for cases 
   (c) and (f))]
and two extreme values of $\ell_{MFP}$ (100 for cases in the left column, i.e.\ $\ell_{MFP}>\sigma$ ,
 and 4 for cases in the right column, i.e.\
  $\ell_{MFP}\ll \sigma$). Spectra are
 reported the $\alpha$ parameter of the corresponding
 statistical analysis. 
 No fluctuations regime is detectable in the more
  diffusive case (right column).} 

 \label{lalande2}
\end{figure}

\begin{figure}[h!]
 \centerline{\scalebox{0.24}{\includegraphics{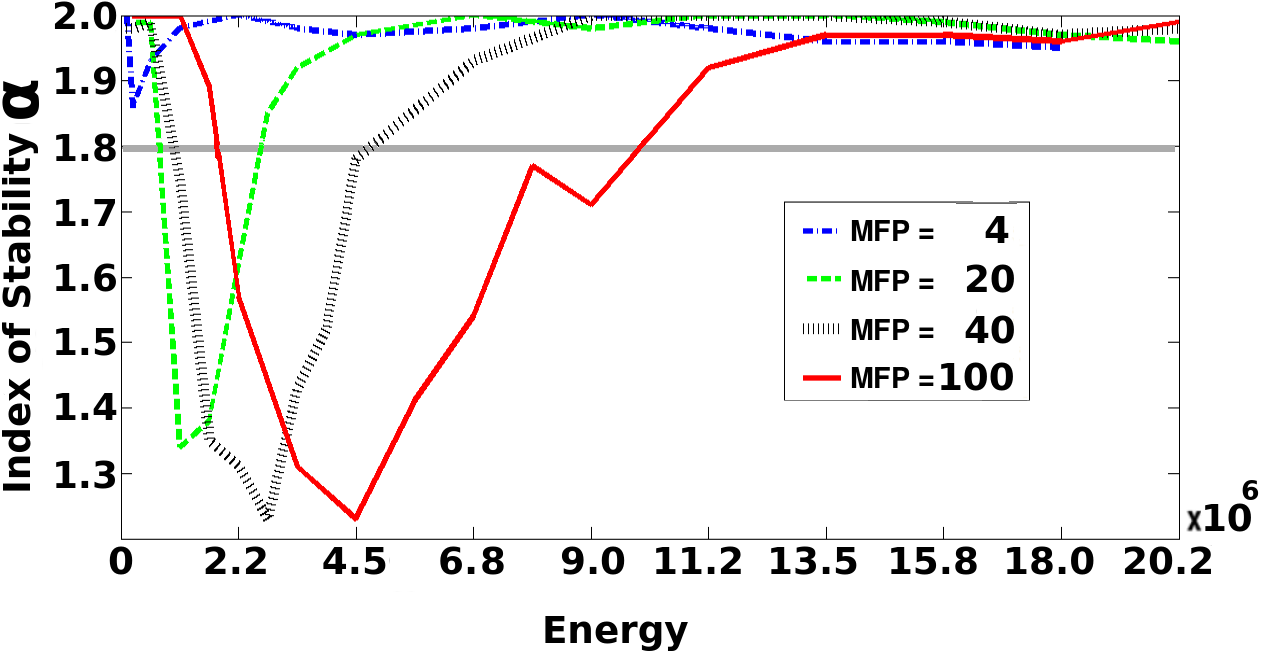}}}
\caption{(Color online) Numerical simulation results for
 the $\alpha$ parameter as a function of pumping
energy (in units of $\hbar\omega_0$) different values of $\ell_{MFP}$, with
  $\sigma = 40$ and $T_p$=0. The
horizontal line at $\alpha=1.8$ is reported as a criterion
 to discriminate the boundary between 
Gaussian and L\'evy regimes (see text).
}
\label{alphanumst}
\end{figure}

\begin{figure}[h!]
 \centerline{\scalebox{0.24}{\includegraphics{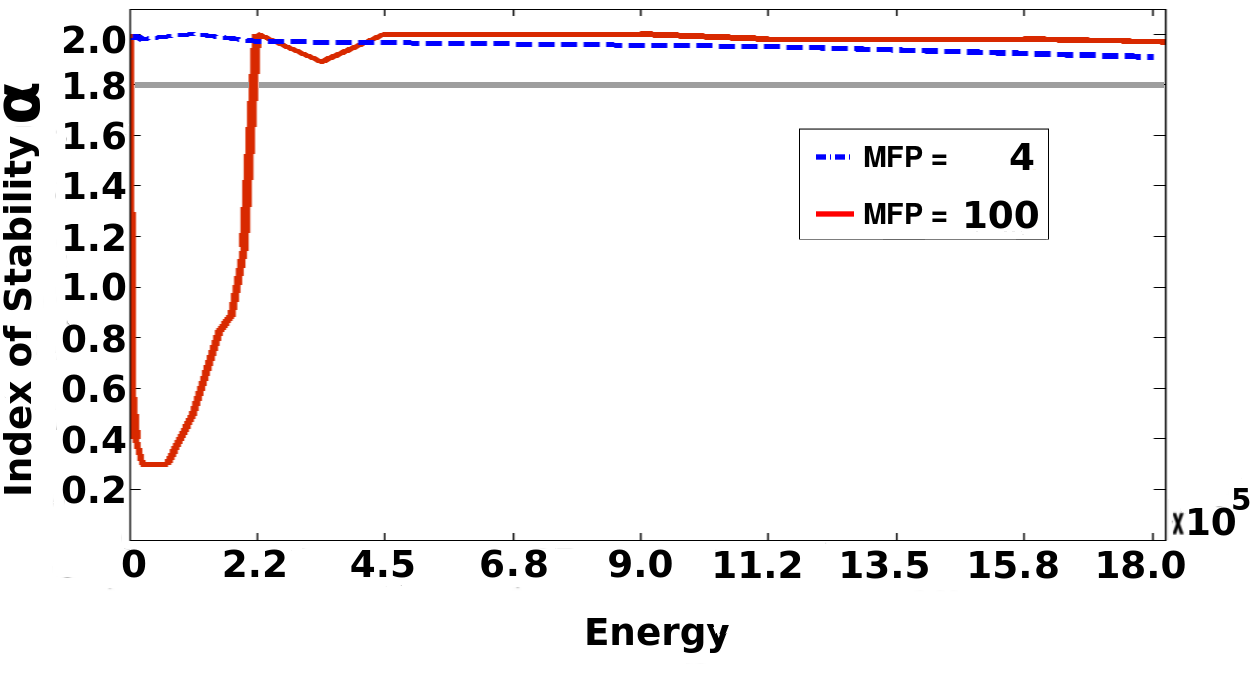}}}
\caption{(Color online) Numerical simulation for a pumping pulse
 with a time duration equal to spontaneous emission lifetime
 ($T_p$ = $\gamma_0^{-1}$ = $10^3$). The plot reports the $\alpha$
 parameter as a function of pumping energy
  for two different $\ell_{MFP}$s, with  $\sigma$ = 40.
 The weakly diffusive active medium ($\ell_{MFP} = 100$)
 is characterized by a broad L\'evy-region with $\alpha$  that
 reaches extreme low values, whereas for strong
 diffusion ($\ell_{MFP} = 4$) $\alpha$ approaches 2 for
 every excitation energy value.}
 \label{final_grad}
\end{figure}

Figure \ref{lalande} shows the simulation results for three
different pump energies. The chosen parameter values
 are $\gamma_0 = 10^{-4}$, a emission linewidth of $w = 50$ channels
 and a $\ell_{MFP}= 40$, whereas the pumping
temporal type is impulsive, with the whole amount of
 energy supplied at the time $t=0$. Some typical
emission spectra are reported in the first column for
 three values of the pumping energies, while in the
second column are reported the histograms of the intensity
 distribution of the spectrum peak as obtained
by 10000 simulations with identical initial conditions.
 The values of the pumping energy are chosen to
be sub threshold, near threshold and above the random laser threshold.
 The threshold is the energy of the change of slope
 of the peak intensity behavior vs pump pulse energy
 averaged on more than 1000 simulations.
  The fluctuations of the spectral intensity at
  different frequencies  near the maximum were
 completely uncorrelated. The characteristic parameters
 of the L\'evy-stable are estimated via a regression-type fitting 
\cite{alpha1,alpha2,alpha3}.

It is important to note that in correspondence of the 
intermediate pump
energy value,  the emission spectrum is characterized 
by the presence of spikes at random frequencies
with strong fluctuations of the peak intensity. 
The fitting of the histogram yields indeed $\alpha=1.43$,
 demonstrating that 
for this parameters choice the system is in the regime of
 L\'evy fluctuations.  
Conversely, for the lower and the
higher pumping levels, fluctuations are modest and $\alpha$ 
is approaching the Gaussian value of 2.
For this simulation $\ell_{MFP}=50$, i.e. comparable to $\sigma=40$.

In Fig. \ref{lalande2}  typical spectra and  two extreme values of
 the $\ell_{MFP}$ are reported:
The data on the left are calculated with a $\ell_{MFP}=100$, that
 is larger than pumped region
dimension ($\sigma$ = 40), and the data on the right with 
$\ell_{MFP}=4$ which is much smaller. For
the same energies are also reported the values of $\alpha$ index. 
It is evident that the spectra are drastically different in
the two cases: In the weakly diffusive case ($\ell_{MFP}=100$) 
the intensity fluctuations are
L\'evy-distributed even at larger energy while in the 
strongly diffusive case 
($\ell_{MFP}=4$) the distribution is always Gaussian.

For a more quantitative analysis, in Fig.~\ref{alphanumst}, 
we report the $\alpha$ index calculated  as
a function of pump energy for four different values of $\ell_{MFP}$.
 Assuming, as discussed above,
the threshold value of
$\alpha=1.8$ for the passage from Gaussian to L\'evy distribution,
 we note that a strongly diffusive
medium displays a Gaussian intensity distribution at any value of the pump
 pulse. In the case of weaker
diffusive media three different statistical regimes are possible increasing
 the pumping energy: The
intensity distribution is Gaussian at low pump energies, becomes
 a L\'evy-type distribution at
increasing pumping energy and then becomes Gaussian again for
 higher pumping energies.  A key parameter
for the statistical behavior is the ratio of  $\sigma$ 
over $\ell_{MFP}$. When $\sigma$ $\leq$
$\ell_{MFP}$  the random paths in the excited part of the medium
 have a low probability of being
spatially overlapped for low or intermediate pumping energies.
 In this condition a first transition of
the statistical distribution from Gaussian to L\'evy is possible,
 corresponding to a transition from a
spectra dominated by spontaneous emission to one where it is possible
 that rare long random paths within
the active medium, with a very large gain, give rise to spikes in the
 spectrum. By increasing the pump
pulse energy, the number of random walkers increases, as does the probability
 of having spatial overlapping and
temporal coincidence. The random walks are coupled by the population
 inversion, with a general
amplification with limited fluctuations,leading to a second transition
 to a Gaussian distribution.  The
line $\alpha=1.8$  is reported in figure for emphasize the crossovers
 between statistical regimes
(Gaussian for $\alpha>$ or L\'evy for $\alpha<1.8$).
 For $\sigma$ $\gg$ $\ell_{MFP}$
($\ell_{MFP}$ = 4), the random paths are strongly coupled
 and the intensity fluctuations are always
limited presenting a Gaussian distribution both in the regime
 of spontaneous or stimulated emission
($\alpha$ always above 1.8).

From a temporal point of view, a critical issue for the 
emergence of L\'evy statistics is
the comparison of two characteristic times: 
the mean scattering time $T_{sc}=dt/Ps$, which is 40 in
the case of Fig. 1 and 100 and 4 in the cases 
reported on Fig. 2 (in units of $dt$), and the
ballistic time span required to cross 
the gain volume $T_{gv}$, which can be estimated as
$\sigma dt/\ell$, which is 40 in units of $dt$ for all cases.
The presence of L\'evy statistics in the whole 
range of pump energy  is possible when $T_{sc}$
becomes of the same order or larger than $T_{gv}$,
 as it  is showed in cases with a mean
scattering time of 40 and 100, while a 
Gaussian statistics is always observed  when
$T_{sc}$ is much smaller than $T_{gv}$ ($T_{sc}$=4).				

Another set of numerical simulations was performed in order to
 clarify the role of the pumping time
duration $T_p$ on the spectrum characteristics. One may expect
 that a larger $T_p$ 
could lead to an averaging mechanism
inhibiting the L\'evy regime of fluctuations. The results reported
 in Fig. \ref{final_grad} show two
different trends of $\alpha$-values as a function of pumping energy
 supplied to the medium in a time
equal to the spontaneous emission lifetime. In the less diffusive
 medium case, a large zone of L\'evy
distribution is present while, in a more diffusive medium,
 the L\'evy region disappears and the $\alpha$
is approaching 2 for any value of the energy. This result
 indicates that the time duration of the
pumping mechanism does not determine the possibility of
 emerging of strong fluctuations while it is
again crucial the condition of the weak or strong diffusive medium.

One can wonder how this model is connected to the concept
 of mode and whether our results can be interpreted
 in that theoretical frame.    The theoretical approach
 followed in this work suggests an interpretation
 of ``open mode'' as possible random path with a certain spatial profile
 within the active medium; under
 this picture, the competition of modes in the gain has a critical
 role, determining the condition of
 different spectral profiles and the statistical regime of fluctuations.
 The condition of modes
 spatially uncoupled, possible in a weakly scattering active medium,
 can lead to strong fluctuations and
 isolated peaks in the spectrum (L\'evy regime) that disappear
 in the high pump energy condition when
 many modes are populated simultaneously. In the case of  strong scattering
 active medium, the modes are always coupled, leading to weak intensity
fluctuations and a smooth spectrum (Gaussian regime): In this
 condition the L\'evy regime disappears for any pump energy.

\section{Experimental Set Up and Samples}\label{Experimental
 Set Up and Samples}
In our experiments we utilized samples made up a solution of 1 mM
 concentration of Rodhamine 6G dye
dissolved in methanol, with different concentrations of a TiO$_2$
 nanopowder, with a mean radius of 300
nm, suspended in the solution to create disorder. The nanopowder
 concentration was varied from 0.1
to 3 mg/cm$^3$ respectively corresponding to scatterer
 concentrations from 2.2 $\times$ 10$^9$ to
6.7 $\times$ 10$^{10}$ cc$^{-1}$,with an independently measured
 $\ell_{MFP}$ that ranged from 4 to 0.13 mm.
The experimental $\ell_{MFP}$ values are estimated by extrapolation from the values obtained by
a measure of intensity attenuation of the ballistic beam at very low scatterer
concentrations (from 0.003 to 0.03 mg/ml), in samples where the single-scattering
condition is well satisfied.

In order to prevent inhomogeneity and clustering of TiO$_2$ nanopowder,
 the samples are subjected to a
ultrasonic bath before and shuffling  by magnetic agitator
 during measurements.
The sample was excited by the second harmonic of a monomode
 Nd:Yag laser ($\lambda$ = 532.8 nm) with a
10 Hz repetition rate and a time duration of 4 ns, a time
 comparable to the spontaneous emission decay
time of Rodhamine 6G.  The pump beam is focused on a spot with diameter
 of 30 $\mu$m and with an energy
up to 360 $\mu$J. 
 The shot-to-shot emission
 from the sample was collected by an
Ocean Optics Spectrometer with spectral resolution of 0.3 nm.
 An optical fiber can collect the 
radiation emitted from the sample at different tilting angles with respect to the 
pump beam direction. The
spectra reported in this paper are relative to an angle of 15$^\circ$. 

To perform the statistical analysis of the emitted intensity, for each
 laser shot we recorded six values
of the spectra, spaced by 0.5 nm, in a small spectral region around
 the average peak wavelength for a
large number of consecutive spectra, up to 2000 for each pumping
 energy interval.  Great care has been
used to monitor the shot-by-shot pump pulse energy to measure
 the emission characteristics as a function
of pump energy limited to a small energy interval with a posteriori
 analysis in order to eliminate the
fluctuations of the laser pump pulse as spurious cause of fluctuations
 in the random laser emission
intensity. The energy intervals is variable from 1 to 15 $\mu$J to
 follow the critical behavior of the
statistical fluctuation characteristics versus pump pulse energy. 

As a further experimental condition during the measurements we check
 also the condition of monomode laser pulse excitation which assures
 a smooth temporal shape. This issue is also important in order
 to discriminate intrinsic emission statistics of the 
random laser emission from strong temporal laser pulse intensity 
fluctuations that are present in a multimode pump laser.

\section{Experimental Results}\label{Experimental Results}

\begin{figure}[h!]
 \centerline{\scalebox{0.45}{\includegraphics{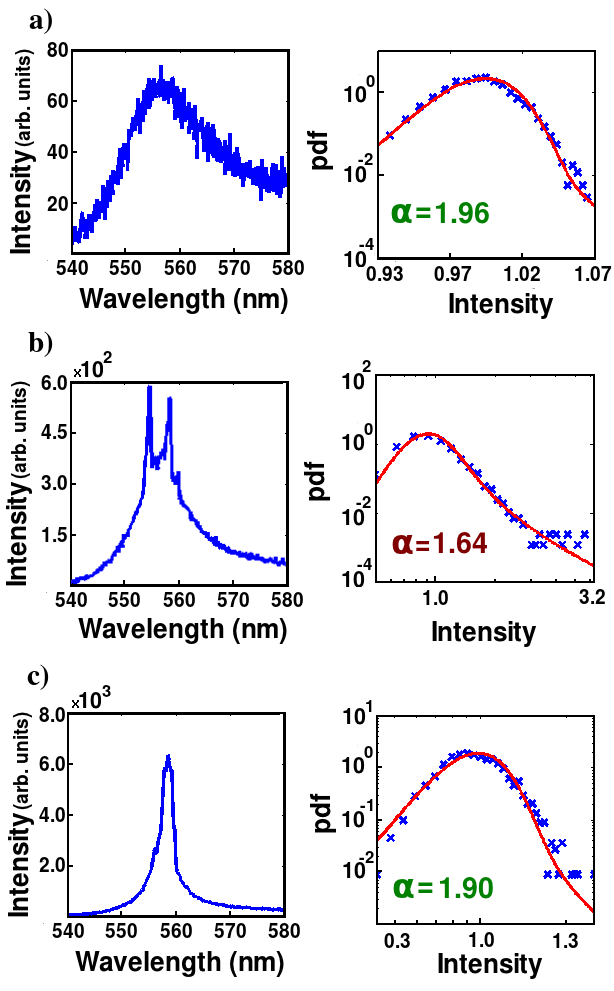}}}
\caption{Experimental results for $\ell_{MFP}$ = 1.3 mm.
 The three values 
of pumping energy are relative to the condition of sub threshold,
 near threshold and far above
 threshold [ 5$\mu$J for case (a), 11$\mu$J for case (b)
  and 396$\mu$J for case (c)].
 On the left column typical spectra are shown, while 
 in the right column
 spectral peak intensity histograms of the experimental
  data and
 relative pdf fits are reported with the value 
 of the $\alpha$ parameter. 
 Intensity distributions are normalized to 
 median and reported in 
logarithmic scale. Similarly the numerical 
simulation shown
  in Fig.\ref{lalande},  in cases (a) and (c) 
  the statistical regime is Gassian, 
  while the case (b) falls in L\'evy 
  fluctuations regime.}  
\label{zone1}
\end{figure}
\begin{figure}[h!]
 \centerline{\scalebox{0.45}{\includegraphics{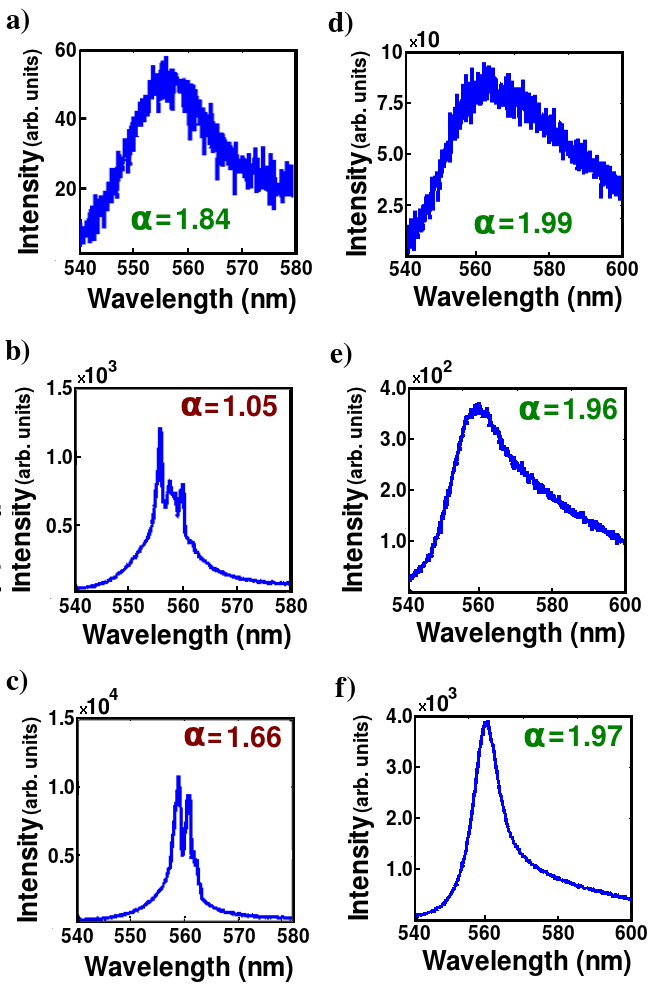}}}
\caption{Typical experimental spectra for 
two values of the $\ell_{MFP}$
 corresponding to stronger and weaker 
 diffusion strength as compared
 to the sample of of figure \ref{zone1} 
 (3.9 mm in cases on the left column
  and 0.4 mm in cases on the right column) 
  for similar values of
 pumping energies (2$\mu$J for case a) and d),
  11$\mu$J for case b) and e) and  360$\mu$J 
  for case c) and f)). In the spectra are 
  reported the $\alpha$ 
parameter of the corresponding statistical 
analysis of the
 experimental data.
 No fluctuations regime is detectable in the more
  diffusive case (right column). } 

 \label{zone2}
\end{figure}

\begin{figure}[h!]
 \centerline{\scalebox{0.3}{\includegraphics{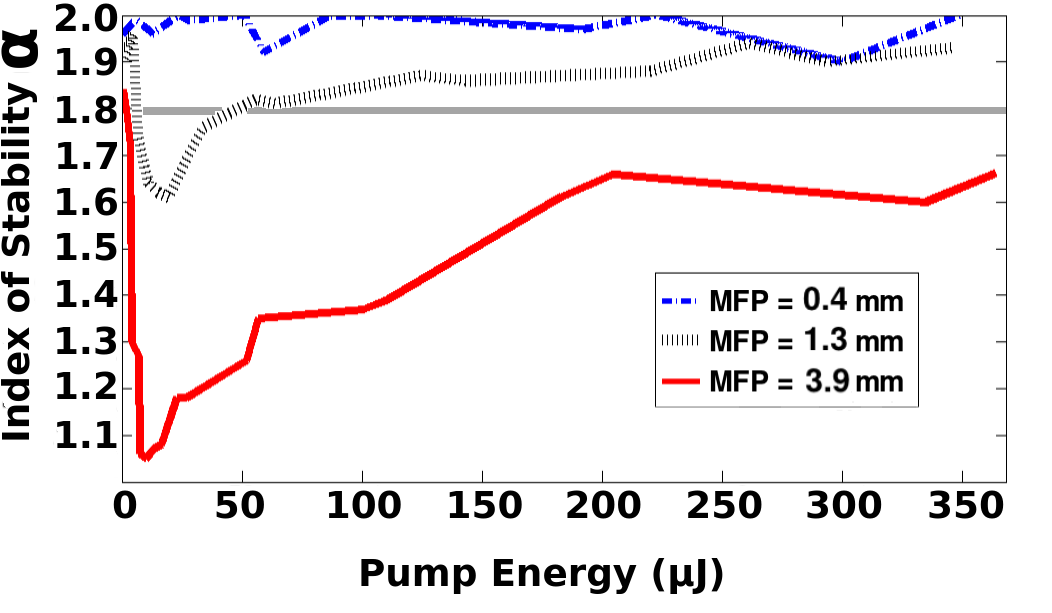}}}
\caption{(Color online) Experimental results for $\alpha$-values
 for three values of the $\ell_{MFP}$, varying the pump energy. }
 \label{alphaexp}
\end{figure}

In Fig. \ref{zone1} typical spectra at the concentration of
 0.3 mg/cm$^3$ (corresponding to a
$\ell_{MFP}$ = 1.3 mm) for three different energies of the pumping 
pulse are reported.  These energies are
chosen to be sub threshold, near threshold and above threshold.  The threshold is
 determined as 
the energy value at which a large change of
slope of the average peak intensity occurs as  a function of
 the pump pulse energy. 

The corresponding intensity histograms are also reported
 with the fit of the  L\'evy-stable 
distribution and corresponding $\alpha$ parameter as done
 for the theoretical computed spectra. The
spectra show that the three different regimes for the intensity
 fluctuation statistics: Gaussian for the
lowest energy ($\alpha$ = 1.96), L\'evy with spiky spectrum
 ($\alpha$ = 1.64) at intermediate energy, and
Gaussian again with a smooth spectrum for the highest 
energy as it was in the numerical simulations. In
Fig. \ref{zone2} are reported spectra for the same three
 energy levels in the cases of 0.1 mg/cm$^3$ and 1
mg/cm$^3$, corresponding to $\ell_{MFP}$ = 3.9 mm and
 $\ell_{MFP}$ = 0.4 mm, respectively. In the case
$\ell_{MFP}$ = 3.9 mm the system display a transition 
from a low-fluctuations regime at low energy  to a
high-fluctuations one for both the other two energy 
values. Is important to note that the value of 360
$\mu J$ is the maximum that we could use to prevent 
damage to the sample, and then a possible
 return to Gaussian statistics for larger pulse energy. 
In the case $\ell_{MFP}$ = 0.4 mm  the spectra are always
 smoothed with a corresponding value of $\alpha<2$ approaching 2.
 The L\'evy region is then absent in this case as in the numerical 
simulations for small $\ell_{MFP}$.  As done in the numerical 
simulation we quantify this behavior in the experimental results
 reporting in Fig. 7 the
$\alpha$-value  versus pump pulse energy for three value 
of the $\ell_{MFP}$. The curve corresponding to  $\ell_{MFP}$ = 1.3 mm 
shows the three different statistical regimes, from Gaussian for low pump
 pulse energy to L\'evy statistics ($\alpha<1.8$) and again Gaussian statistics
 for high pump pulse energy. The more diffusive case, $\ell_{MFP}$ = 0.4 mm,
 presents a closure of the L\'evy regime region with a Gaussian behavior
 throughout the whole energy range. The case of weak diffusion,
 $\ell_{MFP}$ = 3.9 mm, presents a rapid transition to L\'evy regime
 that remains for all the values of the pump pulse energy, and 
 is limited by the constraint of preventing sample damage for greater 
 pumping intensities. The trend of 
$\alpha$ in this last case seems indicate that a transition
 to a Gaussian region could be reached again if a larger energy
 of the pump pulse was achievable without sample damaging.
 The behavior of  the spectra and $\alpha$ value  versus pump
 pulse energy for different values of the $\ell_{MFP}$ show
 common characteristics and the same evolution throughout
 statistical regimes that emerged from the results of our
  theoretical model.

\section{Conclusion}
\label{Conclusion}

We have studied the statistical regimes of a random laser
 performing both simulations and experiment. 
Both the numerical and experimental results consistently
 demonstrate the presence of
three fluctuation regimes upon increasing the pumping
 energy in a weakly diffusive medium. An
initial Gaussian regime is followed by a L\'evy one,
 and Gaussian statistics is recovered again for higher
 pump pulse energy. This behavior is always present 
in the condition of weakly diffusive medium
(small $\ell_{MFP}$).

The situation is different for the strongly diffusive 
case (large $\ell_{MFP}$): Here
the region of L\'evy statistics is not detectable
 and the medium always yield a
Gaussian regime with smooth emission spectra. 

In our study the linkage between theory and experiment
 emerges in the framework of spectral analysis of random lasers,
 in particular regarding the origin of random spikes 
  and statistical regimes crossovers, finding that the role of the $\ell_{MFP}$
 and the dimension of the active medium are crucial. 
 From this study it emerges
 also that the excitation time is not crucial for the statistical
 regime characteristics
The experimental and numerical data in Figs.~\ref{lalande2} 
and \ref{zone2} are theoretically explained 
by the simple model which has been recently been discussed 
in Ref.~\cite{Lepri13}. Such model arises by
a number of simplifying assumptions and amounts to a stochastic 
partial differential equation for the
energy  field with contains a multiplicative random-advection 
term, yielding  intermittency and power-law
distributions of the field itself \cite{Lepri13}.  
The analysis of such equation indicate that
L\'evy-type of fluctuations are more likely  to be observed 
close to threshold and for small 
samples when the  mean free path of photons is not too 
short with respect to the sample size (and more
generally to the size of the gain volume).  In this case the 
random advection term dominates   the
diffusive and gain terms and is responsible for unconventional fluctuations.
  Its relevance is gauged by
the effective noise strength  which in the present case should be
 roughly of the order of 
$(\ell_{MFP}/\sigma)^2$. It is clear that the difference
 observed between samples with  small and large
$\ell_{MFP}$ reported in Figs.~\ref{lalande2} and \ref{zone2} 
 nicely fit with this  prediction.  A more
quantitative comparison will be addressed in the future. 
Such an fact is, however, a promising indication
that connections between the physics of random 
lasers and the theory of nonequilibrium phenomena
in spatially extended systems are likely to be established.

\end{document}